%% file: workshop.tex
\documentclass[
]{ceurart}

\usepackage[show]{chato-notes}

\usepackage[utf8]{inputenc}
\DeclareUnicodeCharacter{2212}{\ensuremath{-}} 

\usepackage{adjustbox}
\usepackage{tabularx}
\usepackage{tikz}
\usetikzlibrary{positioning}
\usepackage{subcaption}

\usepackage{listings}
\newcommand{\para}[1]{\paragraph{\textnormal{\textbf{#1.}}}}

\usepackage{enumitem}
\newcommand{\uls}{\begin{itemize}[leftmargin=*]}
\newcommand{\ule}{\end{itemize}}
\newcommand{\ols}{\begin{enumerate}[leftmargin=*]}
\newcommand{\ole}{\end{enumerate}}
\newcommand{\li}{\item}

\usepackage{pifont}

\newcommand{\CM}[1]{\textcolor{black}{#1}}
\newcommand{\fzw}[1]{\textcolor{black}{#1}}

\DeclareMathAlphabet{\pazocal}{OMS}{zplm}{m}{n}
\DeclareMathAlphabet{\pazobfcal}{OMS}{cmsy}{b}{n}

\begin{document}

\copyrightyear{2022}
\copyrightclause{Copyright for this paper by its authors.
  Use permitted under Creative Commons License Attribution 4.0
  International (CC BY 4.0).}

\conference{ECIR'25: QPP++ Workshop,
  April 06--10, 2025, Lucca, Italy}

\title{Revisiting Query Variants: The Advantage of Retrieval Over Generation of Query Variants for Effective QPP}


\author{Fangzheng Tian}[%
orcid=0009−0000−3282−0220,
email=f.tian.1@research.gla.ac.uk,
]
\address{University of Glasgow, Scotland, United Kingdom}
  
\author{Debasis Ganguly}[%
orcid=0000−0003−0050−7138,
email=debasis.ganguly@glasgow.ac.uk,
]

\author{Craig Macdonald}[%
orcid=0000−0003−3143−279X,
email=craig.macdonald@glasgow.ac.uk,
]

\cortext[1]{Corresponding author.}


\begin{abstract}
Leveraging query variants (QVs), i.e., queries with potentially similar information needs to the target query, has been shown to improve the effectiveness of query performance prediction (QPP) approaches. Existing QV-based QPP methods generate QVs facilitated by either query expansion or non-contextual embeddings, which may introduce topical drifts and hallucinations. In this paper, we propose a method that \CM{retrieves} QVs from a training set (e.g., MS MARCO) for a \CM{given} target query of QPP. To achieve a high recall in retrieving queries with the most similar information needs as the target query from a training set, we extend the directly retrieved QVs (1-hop QVs) by a second retrieval using their denoted relevant documents (which yields 2-hop QVs). Our experiments, conducted on TREC DL'19 and DL'20, show that the QPP methods with QVs retrieved by our method outperform the best-performing existing generated-QV-based QPP approaches by as much as around 20\%, on neural ranking models like MonoT5.
\end{abstract}


\maketitle

\section{Introduction}

Query performance prediction (QPP) is the task of estimating the quality of retrieval results for a \textbf{target query} without relevance judgements. As neural ranking models are increasingly applied in IR, developing effective QPP methods for them has become an important task.
However, the common unsupervised post-retrieval QPP baselines, such as NQC~\citep{NQC}, usually perform poorly for neural rankers where the retrieval score distribution is not suitable for distinguishing retrieval quality based on statistical characteristics~\citep{WRIG}. We argue that the approach of~\citet{ReferenceBasedQPP} that leverages the QPP results of the queries with similar information needs as the target query (i.e. \textbf{query variants}, abbr. QVs), which is useful for improving the QPP baselines' effectiveness, can be potentially suitable for QPP in neural rankers.

To provide reliable signals for incorporating the QPP of the target query, \textbf{keeping the information needs relatively invariant across the QVs is a core task in QV-based QPP~\citep{Oleg_2019}}. The existing methods mainly use feedback or non-contextual embeddings to automatically generate QVs~\citep{WRIG}, which are prone to contain hallucinations and information drifts~\citep{Doc2Query--}.
To tackle this difficulty, in this paper, we propose a novel QV retrieval method from an IR training set. According to our experiments, QPP \CM{using} the QVs retrieved by our proposed methods shows better effectiveness than baselines.

\begin{figure}[h]
\centering
 \includegraphics[width=0.55\columnwidth]{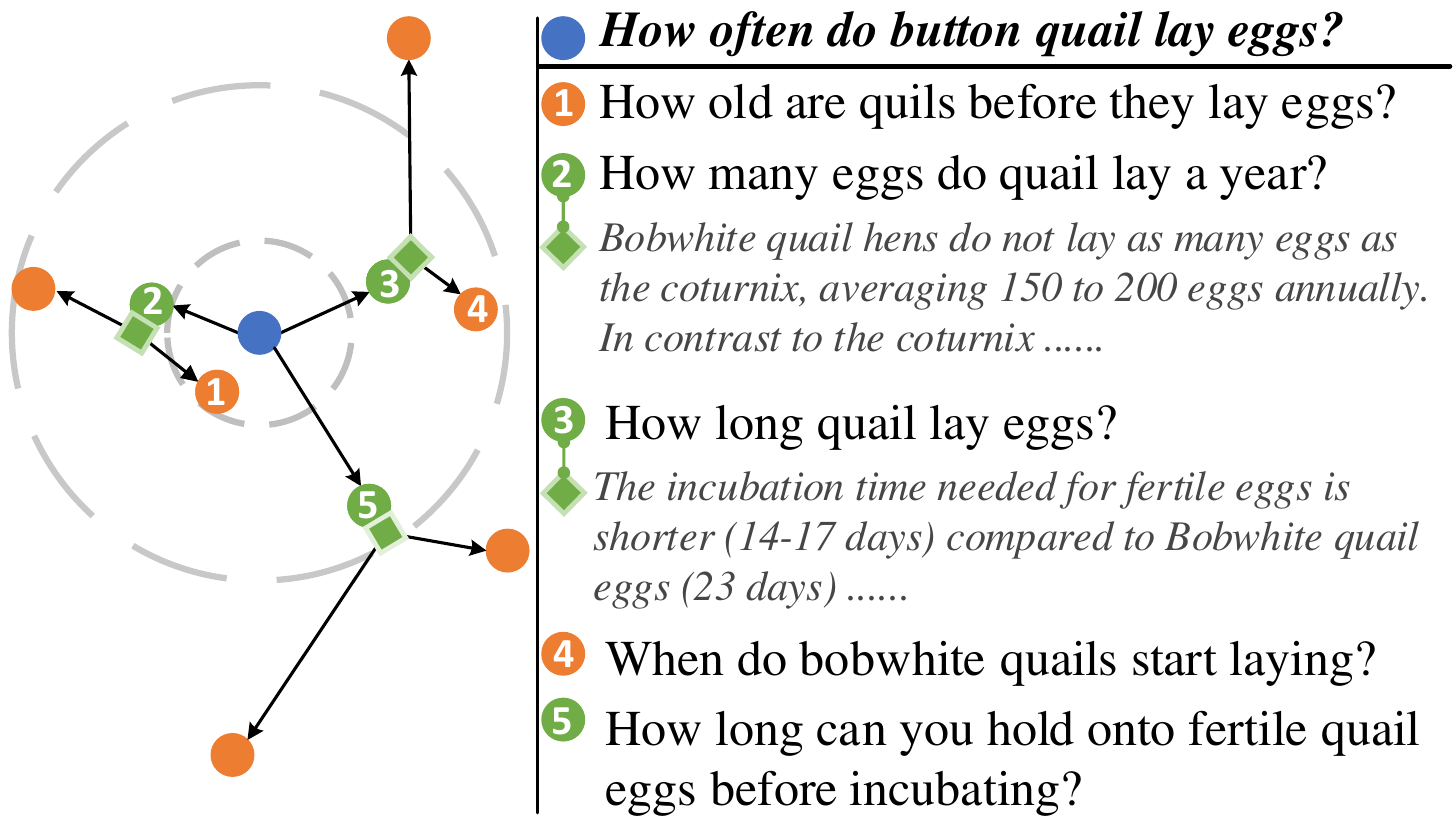}
\caption{An visualisation of the idea behind QPP with \textit{retrieved} QVs. For a target query (the blue point), the first step is to retrieve a list of \textit{1-hop} QVs (the green points) from an index constructed from the training query set. In the second step, the relevant document associated with each \textit{1-hop} candidate (the italicised texts) is used as a query to further retrieve \textit{2-hop} QVs (the amber points). The combination of \textit{1-hop} and \textit{2-hop} QVs comprises the target query's 2-hop neighbourhood, which defines the range of the final candidate QV set.
}
\label{fig:neighbourhood}
\end{figure}

The goal of retrieving QVs from a training set is to find the queries that are most similar to the target one, which will then be included in a candidate QV set for QPP. However, due to the inherent deficiency of representing an information need by a short query~\citep{Oleg_2019}, directly retrieving QVs based on the target query may be incapable of getting a high recall in retrieving similar queries. To address this problem, we propose to use the relevant documents of the directly retrieved QVs (referred to as \textbf{1-hop QVs}), which are available as a part of the MS MARCO training set, to include more candidate QVs (referred to as \textbf{2-hop QVs}). Specifically, the relevant documents of the 1-hop QVs are formulated as queries to retrieve 2-hop QVs from the query index, which is a process similar to relevance feedback~\cite{Lavrenko_RLM2001:RBL:383952.383972}.
Since 2-hop QVs implicitly contain relevant information from the 1-hop QVs, they can still be topically related to the information the target query's information needs and act as useful references in a QV-based QPP model. The idea of the neighbourhood expansion from 1-hop QVs to 2-hop QVs is depicted in Figure~\ref{fig:neighbourhood}.

The candidate QVs within the 2-hop neighbourhood
(illustrated in Figure~\ref{fig:neighbourhood})
comprise the final QV candidate set. In the next step, the candidate QVs are re-ranked according to the similarities between their top retrieved documents and the target query's retrieval result (by means such as RBO~\citep{rbo}), which is a more effective indicator of the similarity between information needs than simply measuring similarity between queries. The top-ranked QVs after the re-ranking are subsequently used in the QV-based QPP.
In our experiments, compared with existing QV-based QPP methods, the QPP estimations \CM{using the QVs retrieved by our proposed} methodology largely enhance the QPP effectiveness for neural retrieval models.
For instance, our QPP method based on retrieved QVs improves the QPP effectiveness on MonoT5 by around 20\% in comparison to that of generated QVs.
Therefore, the innovation in exploiting the resources provided by IR training sets to retrieve QVs that can be helpful for QV-based QPP is the main \textbf{contribution} of our work.

In the next section, we review the work related to QV-based QPP. Section~\ref{sec:methodology} introduces the methodology of QV retrieval. Section~\ref{sec:experiment_settings} is about the settings and results of the experiments which test the QPP approaches based on the QVs retrieved by our proposed method.
Section~\ref{sec:conc} concludes this paper with directions for future work.

\section{Related Work}

\para{Unsupervised QPP methods} 
QPP models can be categorised into pre- and post-retrieval ones based on whether the retrieval results are available to a prediction model. In \textbf{pre-retrieval QPP} methods, a query's performance is estimated by only using the information from the query and the collection statistics, and thus their prediction objective is either the specificity~\citep{NeuralPreQPP} or difficulty~\citep{embWVQPP} of a query. It can be interpreted as the prediction of the expected retrieval quality which is not specific to any retrieval models. In contrast, \textbf{post-retrieval QPP} methods are capable of utilising useful signals from the content and the scores of the top-retrieved documents to predict the retrieval quality~\citep{Clarity}. The underlying hypothesis common to various post-retrieval predictors is the
separability of the top retrieved documents from the remaining ones: a high separability is a likely indicator of the relevance of the top-retrieved documents~\citep{ImproveQPPbyStandardDeviation}. To avoid the effect of outliers, some QPP approaches randomly sample subsets of the top-retrieved documents and then aggregate the predictions over each subset~\citep{uef_kurland_sigir10}.

\para{QV-based QPP methods}
In principle, any statistical estimation method can be improved with the availability of a large number of observation points~\citep{statCompNonDetIRSysUTwoDimVar}. 
In the context of QPP, since a post-retrieval estimator relies on the computation of statistical measures such as the variance of the retrieval score~\citep{NQC}, the estimate for a query can be improved by leveraging information from the queries with information needs similar to the target query (referred to as query variants (QVs) in~\citep{Oleg_2019}).
Usually, an unsupervised QV-based QPP method first generates QVs using relevance feedback~\citep{NasreenJaleel_RM3} or word2vec embeddings~\citep{w2v:mikolov}, \CM{as done by} WRIG~\citep{WRIG}. Then, it uses each generated QV to retrieve \emph{reference lists} of documents for QPP~\citep{ReferenceBasedQPP}. A base predictor, e.g., NQC, is applied to each reference list. The weighted average across those predictions is yielded as the final prediction.

Apart from QV generation, retrieving QVs has also been performed by~\citet{ContextRichQPP} in their supervised QPP method. They concatenate the retrieved QVs and their estimated performances (based on the relevance judgements in the training set) to the target query and its retrieval results. In this way, neighbourhood information is encoded in the target query's expression to help the cross-encoder to predict its performance. However, the training queries usually only have shallow annotated relevance judgments, which leads to the concatenated QVs' performances being unreliable. Therefore, their training data might be unavoidably noisy, which can potentially influence QPP effectiveness. 

\para{Research Gaps} Although~\citet{ContextRichQPP} also retrieve QVs from a training set, they did not deeply investigate how to obtain high-quality QVs. In our research, instead of merely using
1-hop QVs which are directly retrieved by the target query, we explore retrieving 2-hop QVs to extend the candidate QV set to improve the effectiveness of QV-based QPP. The details about our QV retrieval model \CM{follow} in the next section.

\section{Query Variant Retrieval based QPP Methodology}~\label{sec:methodology}

This section first provides a brief technical introduction to the QV-QPP model of~\cite{Oleg_2019} (Section \ref{ss:aggregate_qvs}), following which we describe the proposed QV retrieval component in detail (Section \ref{ss:retrieve_queries}).

\subsection{Smoothing Estimates from QVs}\label{ss:aggregate_qvs}

A post-retrieval QPP model $\phi$ takes a target query $Q$ and its retrieval result $L_M(Q)$ obtained by a \textbf{target retriever} $M$ as input. $\phi$ outputs a score to indicate the quality of $L_M(Q)$. Formally speaking, $\phi: L_M(Q) \mapsto \mathbb{R}$. In QV-based QPP models, the predictor function $\phi$ depends not only on a target query and its retrieved documents but also on a set of query variants (QVs) that contain similar information needs as $Q$. Formally speaking, $\phi^+: L_M(Q), \pazocal{E}(Q) \mapsto \mathbb{R}$, where $\pazocal{E}(Q)$ denotes a set of variants of $Q$. 

We develop our QV-based QPP framework based on the method proposed in~\citep{Oleg_2019}, which aggregates the QPP estimates for $Q$ and its QVs by linear combination. Formally, the predictor in~\citep{Oleg_2019} is defined as

\begin{equation}
\phi^+_{M}(L_M(Q), \pazocal{E}(Q)) =
(1-\lambda)\phi(L_M(Q)) +
\lambda\!\!\!\!\!
\sum_{Q' \in \pazocal{E}(Q)}
\frac{
\sigma(Q, Q') \phi(L_{M}(Q'))
}{\overline{\sigma}(Q)
},
\label{eq:basic-knn-qpp}
\end{equation}

\noindent where $\sigma$ is a function measuring the similarity of information needs between a pair of queries $\sigma(Q, Q')$, and $\overline{\sigma}(Q) = \sum_{Q' \in \pazocal{E}(Q)}\sigma(Q, Q')$, which is the average over all the QVs in $\pazocal{E}(Q)$. Equation~\eqref{eq:basic-knn-qpp} linearly combines the QPP estimations for the target query and the QVs (the same base predictor $\phi$ is used for both). The \textbf{relative importance} of the estimates about QVs is controlled by the parameter $\lambda \in [0,1]$. The QPP about QVs is a weighted average across all the QVs in $\pazocal{E}$, where the similarity measure between a pair of queries $\sigma(Q, Q')$ is used to determine the relative contribution (weight) of each QV $Q'$. \CM{Following~\citet{Oleg_2019}, we use rank-biased overlap (RBO) similarity~\citep{rbo} as $\sigma$, as this QPP estimator puts a higher emphasis on the contribution from the QVs that contains similar information needs as the target query.}

In Equation~\eqref{eq:basic-knn-qpp}, base predictor $\phi$ yields the QPP estimates for QVs based on their retrieval results obtained by the same retriever $M$ as the target query. Our proposed QPP framework generalises Equation~\eqref{eq:basic-knn-qpp} by setting an \textbf{internal retriever} for the QVs: 

\begin{equation}
\phi^+_{M'}(L_M(Q), \pazocal{E}(Q)) =
(1-\lambda)\phi(L_M(Q)) +
\lambda\!\!\!\!\!
\sum_{Q' \in \pazocal{E}(Q)}
\frac{
\sigma(Q, Q') \phi(L_{M'}(Q'))
}{\overline{\sigma}(Q)
},
\label{eq:generic-knn-qpp}
\end{equation}

\noindent where $M'$ is an internal retriever of $\phi^+$, which is allowed to be separate from the target retriever $M$. Therefore, when $M$ is a neural ranker, $M'$ can still be an efficient lexical retriever. Moreover, considering the base predictor $\phi$ (e.g. NQC) can be ineffective for neural retrievers~\citep{WRIG}, this separation can bring additional reliability to the QPP estimation for the target query. In our experiments, we set $M'\equiv BM25$.

\subsection{Retrieving the Query Variants}\label{ss:retrieve_queries}

In our proposed method, QVs in $\pazocal{E}$ are retrieved from the training set. The key motivation for using retrieved QVs, instead of generating them based on relevance feedback or non-contextual embeddings, is that these queries are formulated by real users and, thereby, are more likely to represent real-life information needs and less likely to be hallucinatory.

\para{1-hop QV retrieval}
In the 1-hop QV-based QPP approach, the QV set in Equations \eqref{eq:generic-knn-qpp} corresponds to $\pazocal{E}^1(Q)$, which is a set of top-$k$ most similar queries directly retrieved from a query collection $\pazocal{Q}$ by the target query $Q$. This makes $\pazocal{E}^{1}(Q)$ depending on a similarity function that, given a target query $Q$, is used to compute its $k$ nearest neighbours, which is denoted as
\begin{equation}
\pazocal{E}^{1}(Q) \equiv \pazocal{N}_k(Q, \sigma) \subset \pazocal{Q}, \label{eq:topk_queries}    
\end{equation}
where $\pazocal{N}_k$ is a neighbourhood of size $k$, and $\sigma$ is a similarity function. In practice, we employ RBO as $\sigma$, similar to experiments conducted in~\cite{WRIG,Oleg_2019}. However, since the RBO similarities are the rank-biased overlap between the retrieval results of two queries, it can only be calculated once we have got the retrieval results for them. Therefore, the process of retrieving candidate QVs and then finding the $k$ QVs most similar to $Q$ is like retrieval and re-ranking in retrieving documents.
Firstly, we obtain a candidate set of $n (>k)$ top-retrieved queries using an indexed query collection $\pazocal{Q}$. Then, we compute each candidate's RBO (with respect to the target query) to rerank the list and extract the top-$k$ QVs with the largest RBO scores. This set forms the set of 1-hop QVs. As specific choices for retrieving the 1-hop candidates, we employ \CM{both} a BM25 sparse index and a dense index of SBERT~\cite{SBERT} embeddings.

\input{images/schematic_workshop}

\para{2-hop QV retrieval}
A limitation of retrieving 1-hop QVs is that the queries most similar to $Q$ may not be easily retrieved based on $Q$ itself. The root of this low recall in retrieving high-quality QVs is the ambiguity of representing an information need by a short query. To alleviate this limitation, we leverage the associated relevant documents of each 1-hop query to expand the candidate QV set with 2-hop QVs, which has a bigger chance of retrieving more high-quality QVs. 

Let us denote the set of relevant documents for each query $Q' \in \pazocal{Q}$ by $R(Q')$. We argue that this relevance information for a 1-hop QV $Q'$ can further help to identify the actual information need of $Q'$, which, in turn, is potentially close to that of $Q$. In the light of this idea, we (1) construct a pseudo-query $Q'_D$ from each document $D \in R(Q')$, then (2) execute $Q'_D$ on the index of queries $\pazocal{Q}$ to obtain another list of ``2-hop'' queries $\pazocal{E}(Q'_D)$. The idea is visualised in Figure~\ref{fig:schematic}, where the set of relevant documents for each first-level QV is used to retrieve the second-level ones (the bottom row of queries). Formally,
\begin{equation}
\pazocal{E}^{2}(Q) = \bigcup_{Q' \in \pazocal{N}_k(Q,\sigma)} \bigcup_{D \in R(Q')} \pazocal{N}_k(Q_D,\sigma) \label{eq:2hopqueries}, 
\end{equation}
where $\pazocal{E}^{2}(Q)$ denotes the set of 2-hop QVs. We then merge the $\pazocal{E}^{1}(Q)$ and $\pazocal{E}^{2}(Q)$ as the final candidate QV set from where we select the QVs for QPP:
\begin{equation}
 \pazocal{E}^{+}(Q) = \pazocal{E}^{1}(Q) \cup \pazocal{E}^{2}(Q) \label{eq:final_candidate_set}, 
\end{equation}
where $\pazocal{E}^{+}(Q)$ is the merged QV set containing both 1-hop and 2-hop retrieved QVs of $Q$. Similar to the 1-hop approach, we compute the RBO scores across the QVs in  $\pazocal{E}^{+}(Q)$ and take the QVs with the highest-$k$ RBO scores as the final nearest-$k$ QVs. Those QVs can subsequently used in the QPP framework as described in Equation~\eqref{eq:generic-knn-qpp}.

\section{Evaluation} \label{sec:experiment_settings}

\subsection{Research Questions}
\label{ssec:expsetup:datasets}
The objective of our experiments is to confirm if the QVs retrieved by the proposed method can lead to more effective estimations of retrieval quality.
Therefore, we raise the following research questions about the effectiveness of applying retrieved QVs for QPP and the necessity of using 2-hop QVs.

\uls
\li \textbf{RQ-1}: \textit{Are retrieved query variants from a training set more effective than query variants generated by non-contextual embedding or relevance feedback approaches} (namely the ones used in~\citet{WRIG})?

\li \textbf{RQ-2}: \textit{Can the information from the relevant documents of 1-hop QVs be useful to yield a candidate QV set leading to better QPP estimates}?

\ule

\subsection{Experiment Settings}\label{ssec:expsetup:pipelines}

\para{Datasets} We conduct our experiments on the MS MARCO passage collection benchmark~\citep{MSMARCO-DATASET} with queries from TREC DL'19~\citep{TREC-DL-2019-overview} and DL'20~\citep{TREC-DL-2020-overview} test sets.
The QV candidates are retrieved from the MS MARCO training set containing over 800K examples of query-relevant document pairs.

\para{Target Retrievers} Since the objective of this research is enhancing QPP effectiveness for neural IR, we evaluate our proposed QPP approach and baselines on the following neural ranking models:
(i) \textbf{BM25>>MonoT5}~\citep{monot5}~\footnote{Huggingface id {\tt castorini/monot5-base-msmarco}}: After retrieving an initial candidate of 100 documents with BM25 (using a sparse index), we re-rank the top 100 results by MonoT5, a cross-encoder mode.
(ii) \textbf{BM25>>BERT}~\citep{PARADE}\footnote{Huggingface id {\tt capreolus/bert-base-msmarco}}: BERT is a contextualised language model that can be used to obtain embeddings of queries and documents utilising the classification token (CLS)~\citep{BERT}. We use a model that was fine-tuned on the MS MARCO Passage dataset to re-rank the top 100 BM25-retrieved passages.
(iii) \textbf{TCT-ColBERT}\citep{TCT-ColBERT}: A neural retriever trained by distilling knowledge from ColBERT~\citep{ColBERT}. We use this model to test the performance of our QPP methods for end-to-end neural retrieval.

\para{Internal Retriever (Retrieval of Documents with QVs)}
As described in Section~\ref{ss:aggregate_qvs}, the internal retriever $M'$ in Equation~\eqref{eq:generic-knn-qpp} (i.e., the model which is used to retrieve for the QVs) is always BM25 in our experiments.

\para{Related Query (QV) Retrieval Models} QVs are retrieved from the MS MARCO training query set by: (i) \textbf{BM25}: A term-weighting, document length normalisation-based classic IR model. (ii) \textbf{Sentence-BERT}~\citep{SBERT}: A CLS-pooling-based bi-encoder BERT model pre-trained on the MS MARCO training set triples, which is abbreviated as \textbf{SBERT} in the rest of this paper.
For all the QV-based QPP methods, we re-rank the candidate QVs by their \textbf{RBO}~\citep{rbo} similarities with the target query and then take the top-$k$ ones for the smoothing estimation in Equation~\eqref{eq:generic-knn-qpp}.

\para{Parameters and Settings}
We employ \textbf{Kendall's $\tau$} as the QPP effectiveness evaluation metric.
Since different target IR measures can cause significant variations in QPP outcomes \cite{AnalyseVariantInQPP}, we report our experiments with two target IR metrics, namely AP@100 and nDCG@10 - the latter being top-rank precision-focused.
For computing AP, documents with a relevance label of 2 or higher are considered relevant~\cite{TREC-DL-2019-overview}.

We evaluate the experimented QPP methods with a 2-fold setup: in the first fold, hyper-parameters are trained on DL'19 and tested on DL'20, and the other way around in the second fold. The hyper-parameters trained in every QV-based QPP method (named with `QV*') are: (i) $k$, the number of QVs used in the smoothing estimation, and (ii) $\lambda$, the relative importance of the QV-based QPP smoothing (see Equation \eqref{eq:generic-knn-qpp} for more details).

\subsection{Proposed QPP Methods and Baselines}\label{baselines}
As a common naming convention, we use the common prefix `QV' to name the QV-based QPP approaches.
As QPP baselines, we investigate the following approaches:

\uls

\li \textbf{Predictors without QVs:} (i) \textbf{NQC}~\citep{NQC}, which estimates the retrieval quality by measuring the skewness of retrieval score distribution in the top-ranked documents, and (ii) \textbf{UEF}~\citep{uef_kurland_sigir10}, which samples subsets from the top-ranked documents for estimating the robustness of retrieval quality. We choose NQC as the base predictor for UEF, as \CM{described} on~\citep{WRIG}. Both NQC and UEF also serve as base predictors for the QV-based methods.

\li \textbf{Predictors using generated QVs:}
We employ the methods proposed in~\citep{Oleg_2019}, where the QVs (set $\pazocal{E}$ of Equation \eqref{eq:generic-knn-qpp}) are generated by the methodology proposed in~\citep{WRIG}. There are two methods with different ways of QV generation: (i) \textbf{QV-RLM}: generating QVs by query expansion via a Relevant Language Model (RLM), which is RM3~\citep{NasreenJaleel_RM3} in our implementation, (ii) \textbf{QV-W2V}: generating QV by adding semantically similar terms by using the non-contextual embedding space of Word2Vec~\citep{w2v:mikolov}.

\li \textbf{Supervised Predictors:}
To compare our QV-based unsupervised QPP methods with their supervised counterpart, we implement the cross-encoder-based QPP method \textbf{BERTQPP}~\citep{qppbert}. We also implement a variant of it, \textbf{BERTQPP-QV}~\citep{ContextRichQPP}, which considers QVs and their estimated query performance when encoding the target query.

\ule

\para{Experimented proposed predictors with retrieved QVs}
The key difference between our methods and the QV-based baselines (QV-RLM and QV-W2V) is the usage of retrieved QVs from a training set.
We place an `R' in the names of the methods with retrieved QVs following `QV'. Additionally, we set the superscript of `R' as `1' or `2' to distinguish between the methods with 1-hop or 2-hop neighbourhood considered when retrieving the QVs. According to whether the query retriever is the lexical BM25 or semantic SBERT, we test the following four variants of the proposed method:
(1) \textbf{QV-$R^{1}$-BM25}: This method uses 1-hop query variants only, i.e., only uses the set $\pazocal{E}$ of 1-hop neighbouring queries (see Equation~\eqref{eq:topk_queries}) to smooth the predictor. The 1-hop QVs are retrieved by BM25.
(ii) \textbf{QV-$R^{2}$-BM25}: Like QV-$R^{1}$-BM25, it also uses BM25 as the query retriever. However, it extends the candidate QV set with 2-hop QV selection as shown in Equation \eqref{eq:final_candidate_set} and Figure~\ref{fig:schematic}. The superscript `2' on `R' indicates 2-hop QVs are used in this method.
(iii) \textbf{QV-$R^{1}$-SBERT}: This is similar to QV-$R^{1}$-BM25 but employs an embedding-space (semantic) approach to retrieve candidate 1-hop QVs. In particular, we use SBERT\footnote{Huggingface id \texttt{sentence-transformers/all-MiniLM-L6-v2}} as the query retriever.
(iv) \textbf{QV-$R^{2}$-SBERT}: This method employs the same query retriever as QV-$R^{1}$-SBERT. The difference is that it additionally adds 2-hop neighbours into the candidate QV set.
Above retrieval pipelines are implemented by PyTerrier~\citep{pyterrier}

\input{tabdefs/ret_only_main}

\subsection{Main Observations} \label{sec:results}

We first present the main findings of our approach (relative to the baselines) with a single query variant, i.e., we use $k=1$ in the general QV-QPP framework (Equation \eqref{eq:generic-knn-qpp}). The performances of our QPP approach over a range of different numbers of $k$ are reported in \ref{ss:sensitivity}. 
Table \ref{table:ret_qv_main} shows a comparison between the baselines and the proposed approaches when $k=1$. In this table, the two groups of rows correspond to the two experimented base predictors (denoted as $\phi$) - NQC and UEF. Each row in this table is either a baseline or a variant of the proposed retrieved-QV-based method. The results of supervised BERT-QV are listed at the bottom of this table for comparison.
We highlight the following observations to answer RQ1 and RQ2.

\para{RQ1: Retrieved QVs yield better QPP effectiveness compared to RLM and W2V-based generated QVs} In general, the baseline QPP methods with W2V- or RLM-generated QVs, are more effective than the respective base predictors and the supervised predictors, especially for BM25>>MonoT5 and BM25>>BERT where the retrieval scores are bounded by $[0, 1]$. For example, when the retriever is BM25>>MonoT5, compare the $\tau$ values for the AP@100 with NQC (0.1673) vs. QV-RLM (0.3308) in Table \ref{table:ret_qv_main} (this also conforms to the observations reported in~\citep{WRIG}).
Our proposed retrieval-based QV approach outperforms the ones generated by either RLM and W2V. The difference between the proposed methods and QV-RLM (the best baseline method) is larger when the target metric is AP@100, e.g., with NQC as the base predictor, QV-$R^1$-BM25 achieves a $\tau$ of 0.3669 outperforming QV-RLM (0.3045).
The QV-$R^1$-BM25 result is significantly better than NQC (Fisher's $z$ test at a 99\% confidence interval).

\para{RQ2: Utilising relevant information of 1-hop QVs to retrieve 2-hop QVs enhances QPP effectiveness}
This is observed by comparing the results between these two pairs of QPP methods - 1-hop and 2-hop for each QV retrieval method. For instance, Table \ref{table:ret_qv_main} shows that with NQC as the base predictor, the 2-hop method QV-$R^2$-SBERT outperforms the 1-hop method QV-$R^1$-SBERT ($0.4033$ vs.\ $0.3361$). \fzw{This improvement in correlation is at 80\% confidence interval by Fisher's $z$ test.}

\subsection{Parameter Sensivity} \label{ss:sensitivity}

\begin{figure}[t]
\centering
\begin{subfigure}[b]{0.28\textwidth}
 \centering
 \includegraphics[width=\textwidth]{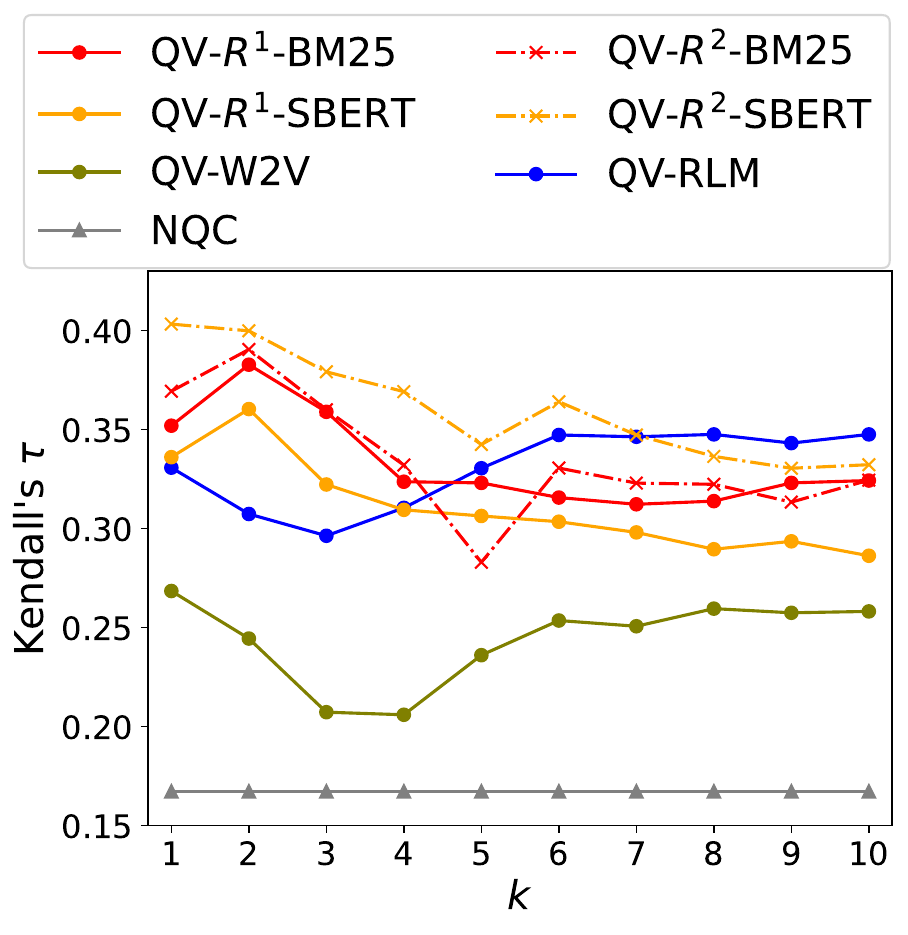}
 \caption{NQC on AP@100}
 \label{fig:apnqc}
\end{subfigure}
\begin{subfigure}[b]{0.28\textwidth}
 \centering
 \includegraphics[width=\textwidth]{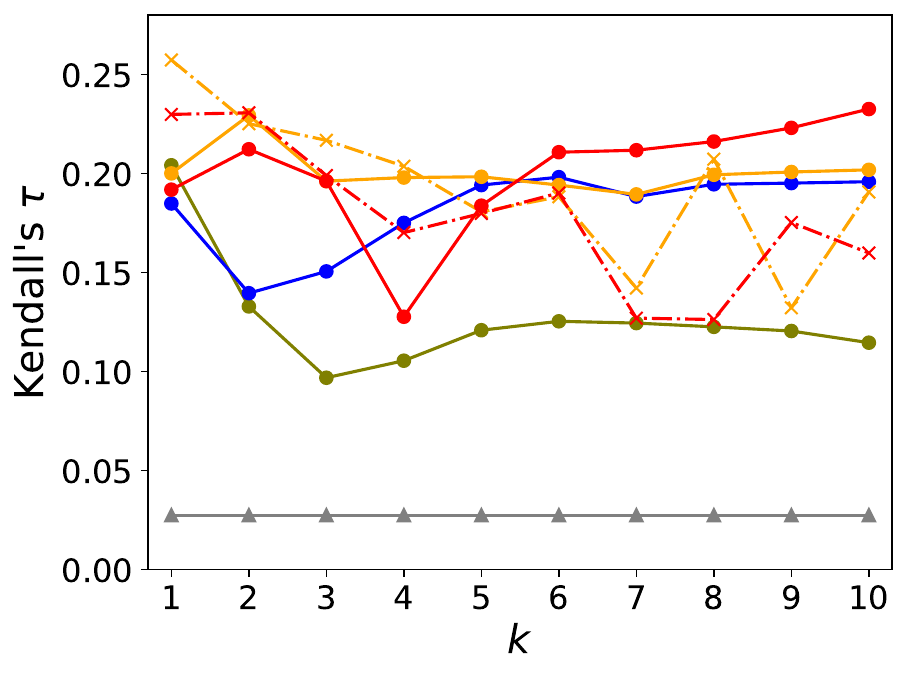}
 \caption{NQC on nDCG@10}
 \label{fig:ndcgnqc}
\end{subfigure}
\caption{A complete spectrum of the results of Table~\ref{table:ret_qv_main} showing the effect of $k$ for each tested QPP method. The base predictor is NQC.
}
\label{fig:mt5_full_k}
\end{figure}

We conduct experiments to investigate the relation between the two hyper-parameters: $k$ - the number of QVs, and $\lambda$ - the relative importance of the QVs in the final QPP estimate described in Equation \eqref{eq:generic-knn-qpp}.

\para{Effect of varying $k$ w.r.t. $\lambda$s in QPP based on retrieved QVs}~\label{sss:sensitivity-k-ret-only}
Figure~\ref{fig:mt5_full_k} shows the QV-based QPP results obtained with $k$ values in $[1, 10]$. \CM{Similar to} Table \ref{table:ret_qv_main}, the values of $\lambda$ are optimised on the train splits of the 2-fold setup.
The general trend observed in all the configurations of the retrieved QVs is that QPP effectiveness decreases with the number of QVs $k$. This is expected because as $k$ increases, QVs with smaller RBO values will be used for QPP. The low-RBO QVs are more likely than the higher-ranked QVs to include information, which is unrelated to the intent of the input query, and hence degrading the QPP effectiveness.

\begin{figure}[h]
     \centering
     \begin{subfigure}[b]{0.28\textwidth}
         \centering
         \includegraphics[width=\textwidth]{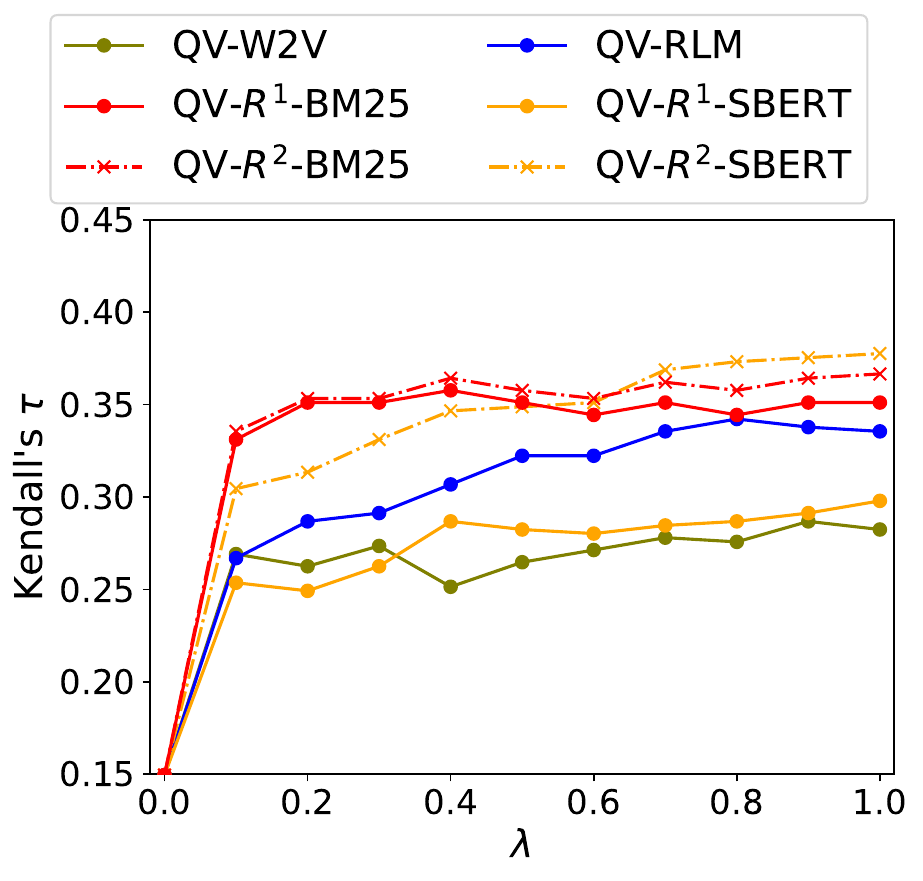}
         \caption{DL'19: $k=1$}
         \label{fig:apnqc_dl19}
     \end{subfigure}
     \begin{subfigure}[b]{0.28\textwidth}
         \centering
         \includegraphics[width=\textwidth]{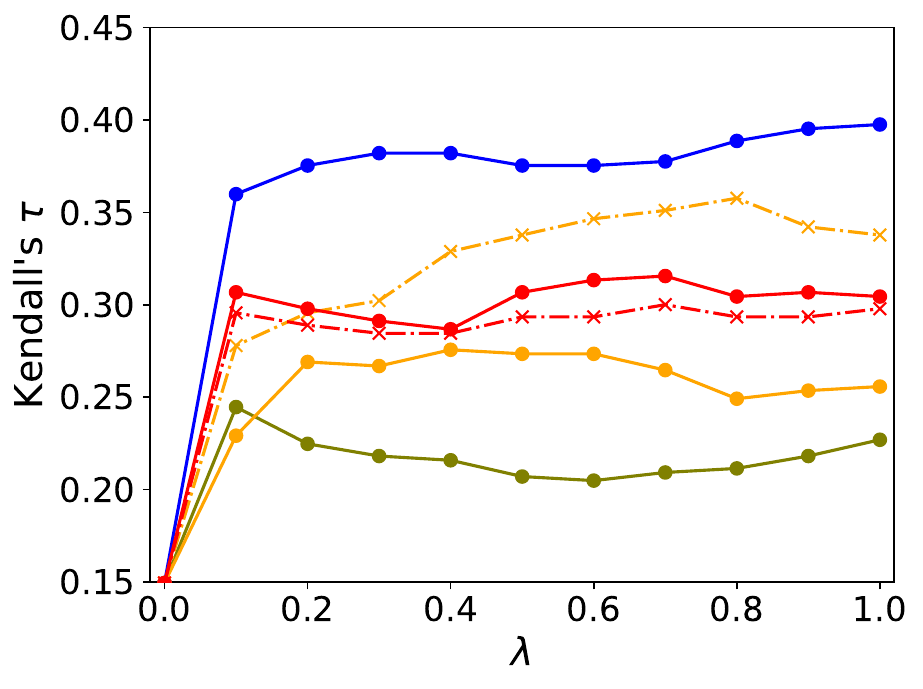}
         \caption{DL'19: $k=3$}
         \label{fig:apuefJ_K3_DL19}
     \end{subfigure}
     \begin{subfigure}[b]{0.28\textwidth}
         \centering
         \includegraphics[width=\textwidth]{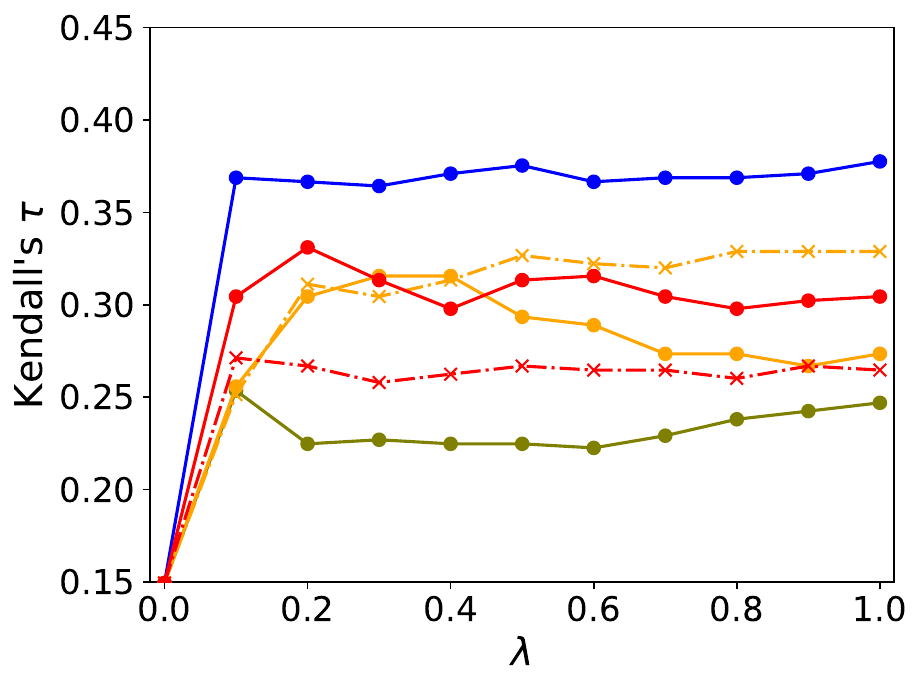}
         \caption{DL'19: $k=5$}
         \label{fig:apuefJ_K5_DL19}
     \end{subfigure}
     \begin{subfigure}[b]{0.28\textwidth}
         \centering
         \includegraphics[width=\textwidth]{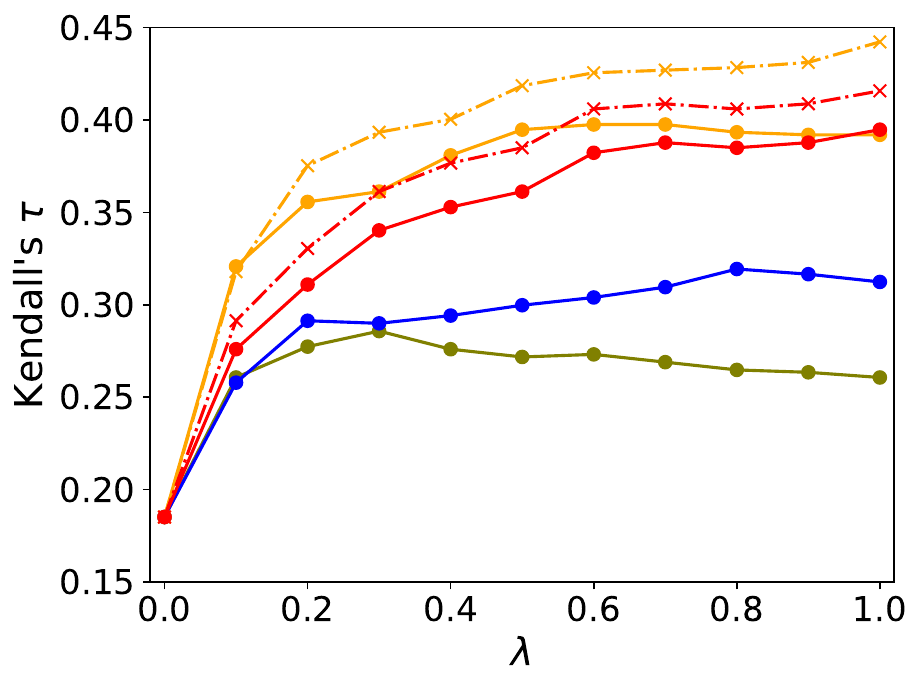}
         \caption{DL'20: $k=1$}
         \label{fig:apnqcJ_DL20}
     \end{subfigure}
     \begin{subfigure}[b]{0.28\textwidth}
         \centering
         \includegraphics[width=\textwidth]{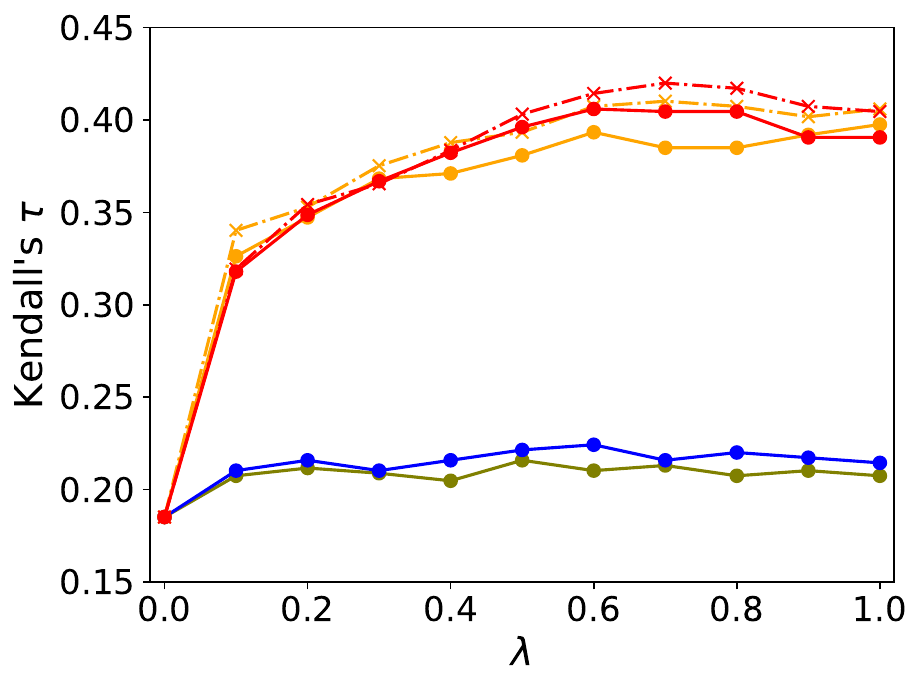}
         \caption{DL'20: $k=3$}
         \label{fig:apuefJ_K3_DL20}
     \end{subfigure}
     \begin{subfigure}[b]{0.28\textwidth}
         \centering
         \includegraphics[width=\textwidth]{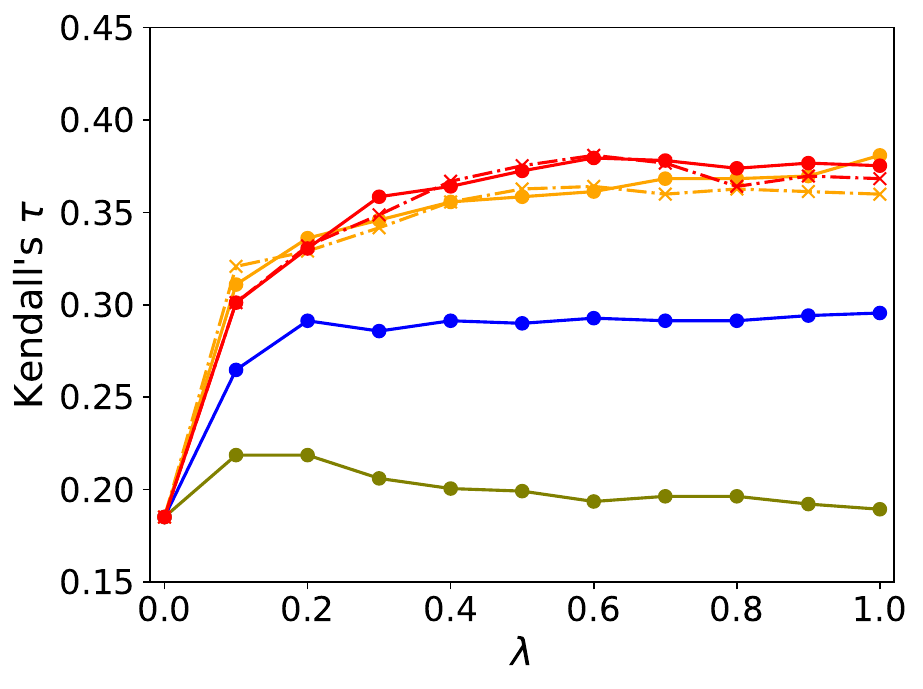}
         \caption{DL'20: $k=5$}
         \label{fig:apuefJ_K5_DL20}
    \end{subfigure}
\caption{Effect of jointly varying both $k$ (\#QVs), and $\lambda$ (relative importance of the QVs) on DL'19 and DL'20 test queries. The base estimator is NQC, and the target IR metric is AP@100.}
\label{fig:lambda_graphs}
\end{figure}

\para{Effect of varying both $k$ and $\lambda$ in QPP based on retrieved QVs}~\label{sss:k_lambda_sensitivity}
Figure~\ref{fig:lambda_graphs} shows the joint effect of both the parameters $k$ and $\lambda$ on the retrieved-QV-based QPP models. To analyse $k$ and $\lambda$ separately, these results are not the average of a 2-fold setup with hyper-parameter tuning; instead, they show the QPP effectiveness of each combination of $k$ and $\lambda$ ($k\in \{1,2,3\}, \lambda\in\{0,0.1,...,0.9,1.0\}$).

First, as a general trend, comparing the proposed methods with QV-W2V and QV-RLM in Figure~\ref{fig:lambda_graphs}, that QPP approaches with retrieved QVs are mostly better than the generated-QV-based QPP methods. The only exception appears in TREC DL'19, where the QPP with the top-3 RLM-generated QVs produces better results than the proposed QPP methods. In all other cases, the proposed methods perform better. Moreover, the QPP methods with 2-hop QVs are consistently better than the ones with 1-hop QVs, thus demonstrating the usefulness of the relevant documents in the training set.

To see what values of $k$ work well, we can compare the plots in Figure~\ref{fig:lambda_graphs} horizontally (e.g., a->b->c). For both DL'19 and DL'20 queries, we see that small $k$s work the best and $k=1$ yields the optimal results.
However, there is a considerable difference in the optimal value of $\lambda$ in DL'19 and DL'20, which can be seen from the fact that the best values for DL'19 usually appear when $\lambda$ is close to 0.3, whereas for DL'20, the highest values are concentrated on the right half of the $\lambda$-axis, where $\lambda>0.5$. 

\section{Concluding Remarks}~\label{sec:conc}

Different from existing QV-based QPP methods, which use manually formulated QVs or generated QVs based on relevance feedback or non-contextual embeddings, we explore the possibility of retrieving QVs within the 2-hop neighbourhood of the target query from a training query set.
We propose to incorporate the retrieved QVs in the QPP for the target query and verify that our proposed methods are more effective than existing QV-based QPP approaches for neural rankers, achieving about $20\%$ increase in AP-$\tau$ relative to the best-performing baseline.

A likely reason for this improvement is that the proposed 2-hop QV retrieval can effectively retrieve the queries which are most similar to the target query in terms of the information need from a training set. For this reason, our work provides an appealing option for improving QPP effectiveness in real-life scenarios where a query log with user interaction history is available.
However, we notice that not every target query can find its query variants from a training set with a fixed number of queries. Therefore, in future, we would like to explore the possibility of integrating QV retrieval with LLM-based QV generation in order to obtain QVs with small topical drifts for every target query.

{
\small

\input{main.bbl}
}

\end{document}

%% file: images/schematic_workshop.tex
\begin{figure}[t]
\centering
\begin{adjustbox}{width=0.55\columnwidth}
\begin{tikzpicture}[
roundnode/.style={circle, draw=black, very thick},
ellipsisnode/.style={circle},
textnode/.style={circle},
squarednode/.style={rectangle, draw=black, fill=black!10, very thick,minimum size=.1\columnwidth},
]
\node[roundnode](Q) {$Q$};
\node[roundnode](vQ1) [below left=8mm and 12mm of Q] {$Q'_1$};
\node[ellipsisnode](vQi) [below=8mm of Q] {$\ldots$};
\node[roundnode](vQn) [below right=8mm and 12mm of Q] {$Q'_n$};
\node[textnode](varQ) [right=2mm of vQn, text width=35mm] {$\pazocal{E}^1(Q)=\bigcup_{i=1} Q'_i$
1-hop neighbors};
\node[squarednode](rQ1) [below=5mm of vQ1] {$R(Q'_1)$};
\node[ellipsisnode](rQi) [below=5mm of vQi] {$\ldots$};
\node[squarednode](rQn) [below =5mm of vQn] {$R(Q'_n)$};
\node[textnode](relset) [right=2mm of rQn, text width=35mm] {$\bigcup_{Q' \in \pazocal{E}^1_k(Q)} R(Q')$
Rel docs of 1-hop neighbors};
\node[roundnode](v2Q11) [below left=5mm and .5mm of rQ1] {$Q''_{1,1}$};
\node[ellipsisnode](v2Qi1) [below=5mm of rQ1] {$\ldots$};
\node[roundnode](v2Qn1) [below right=5mm and .5mm of rQ1] {$Q''_{1,n}$};
\node[roundnode](v2Q1n) [below left=5mm and .5mm of rQn] {$Q''_{n,1}$};
\node[ellipsisnode](v2Qin) [below=5mm of rQn] {$\ldots$};
\node[roundnode](v2Qnn) [below right=5mm and .5mm of rQn] {$Q''_{n,n}$};
\node[textnode](hop2text) [right=2mm of v2Qnn, text width=35mm] {$\pazocal{E}^2(Q) = \bigcup_{i,j} Q''_{i,j}$\\
2-hop neighbors};
%
\draw[->] (Q.south) -- (vQ1.north);
\draw[->] (Q.south) -- (vQn.north);
\draw[->] (vQ1.south) -- (rQ1.north);
\draw[->] (vQn.south) -- (rQn.north);
\draw[->,double distance=3pt] (rQ1.south) -- (v2Qi1.north);
\draw[->,double distance=3pt] (rQn.south) -- (v2Qin.north);
\end{tikzpicture}
\end{adjustbox}\vspace{-1em}
\caption{Schematic representation of the idea of collecting second-hop neighbourhood queries by extending the set of 1-hop neighbours $\pazocal{E}^1(Q)$ with the set of relevant documents $R(Q')$ for each first-hop query $Q'$.}
\label{fig:schematic}
\end{figure}

%% file: tabdefs/ret_only_main.tex

\begin{table}[t]
\centering
\caption{Comparisons between our proposed QPP method (`Ours') and the baselines (`BL') along with the two supervised methods (`Supervised'). QPP effectiveness is measured by Kendall's $\tau$ with AP@100 and nDCG@10 as the target metrics. Reported Kendall's $\tau$ values are averaged over a 2-fold train-test split on TREC DL'19 and DL'20, with $\lambda$ is tunable while $k$ is fixed to 1, as described in Equation \eqref{eq:generic-knn-qpp}. The best results for each metric with a particular base QPP model are underlined. Comparing the best results obtained with NQC and UEF as the base QPP model, the better ones are bold-faced. A `*' alongside one of our proposed QPP approaches over a target metric indicates that the reported result is statistically significant (by Fisher's $z$-test at 99\% confidence level) over the corresponding base predictor.
}
\small
\begin{tabular}{@{}lll ll ll ll@{}}
\toprule
& & & \multicolumn{2}{c}{$\text{BM25$\gg$MonoT5}$}&
\multicolumn{2}{c}{$\text{BM25$\gg$BERT}$}&
\multicolumn{2}{c}{$\text{TCT-ColBERT}$} \\
\cmidrule(r){4-5} \cmidrule(r){6-7} \cmidrule(r){8-9}
$\phi$ & Type & Method & AP-$\tau$ & nDCG-$\tau$ & AP-$\tau$ & nDCG-$\tau$ & AP-$\tau$ & nDCG-$\tau$\\
\midrule 
\multirow{7}{*}{\rotatebox[origin=c]{0}{NQC}} & \multirow{3}{*}{\rotatebox[origin=c]{0}{BL}} 
   & NQC & 0.1673 & 0.0274 & 0.1278 & 0.0391 & 0.3991 & 0.2618\\
 & & QV-W2V & 0.2685 & 0.2041 & 0.2395 & 0.1520 & 0.3923 & 0.2521\\
 & & QV-RLM & 0.3308 & 0.1848 & 0.3045 & 0.1460 & 0.3920 & \underline{0.2848}\\
\cmidrule(r){3-9}

 & \multirow{4}{*}{\rotatebox[origin=c]{0}{Ours}} 
   & QV-$R^1$-BM25 & 0.3520 & 0.1918 & 0.3669* & 0.2081 & 0.3954 & 0.2611\\
 & & QV-$R^1$-SBERT & 0.3361 & 0.2000 & 0.3558* & 0.2016 & \underline{\textbf{0.4177}} & 0.2561\\ 
 & & QV-$R^2$-BM25 & 0.3694* & 0.2298 & 0.3678* & 0.2111 & 0.3878 & 0.2539\\ 
 & & QV-$R^2$-SBERT & \textbf{\underline{0.4033}}$^*$ & \textbf{\underline{0.2573}}$^*$ & \underline{0.4250}* & \underline{0.2517}* & 0.4022 & 0.2614 \\

\cmidrule(l){2-9}

\multirow{7}{*}{\rotatebox[origin=c]{0}{UEF}} & \multirow{3}{*}{\rotatebox[origin=c]{0}{BL}} 
   & UEF & 0.1679 & 0.0341 & 0.1323 & 0.0399 & 0.3908 & 0.2732\\ 
 & & QV-W2V & 0.2674 & 0.1969 & 0.2290 & 0.1583 & 0.3892 & 0.2718\\
 & & QV-RLM & 0.3247 & 0.1940 & 0.2846 & 0.1594 & 0.3845 & \underline{\textbf{0.2998}}\\
\cmidrule(r){3-9}
 
 & \multirow{4}{*}{\rotatebox[origin=c]{0}{Ours}} 
   & QV-$R^1$-BM25 & 0.3438 & 0.1963 & 0.3567* & 0.2247 & 0.3701 & 0.2782\\
 & & QV-$R^1$-SBERT & 0.3481 & 0.2044 & 0.3791* & 0.2159 & \underline{0.4080} & 0.2732\\
 & & QV-$R^2$-BM25 & 0.3597 & 0.2325 & 0.3621* & 0.2079 & 0.3832 & 0.2801\\ 
 & & QV-$R^2$-SBERT & \underline{0.4026}$^*$ & \underline{0.2540}$^*$ & \underline{\textbf{0.4375}}* & \textbf{\underline{0.2630}}* & 0.3619 & 0.2633\\

\midrule
\rowcolor{lightgray}
\multicolumn{2}{c}{Supervised} & BERTQPP & 0.2277 & 0.1746 & 0.1827 & 0.1328 & 0.2238 & 0.1326 \\
\rowcolor{lightgray}
\multicolumn{2}{c}{} & BERTQPP-QV & 0.2432 & 0.1728 & 0.2051 & 0.1459 & 0.2529 & 0.1514 \\

\bottomrule
\end{tabular}

\label{table:ret_qv_main}
\end{table}

%% file: workshop.bbl
\begin{thebibliography}{28}
\expandafter\ifx\csname natexlab\endcsname\relax\def\natexlab#1{#1}\fi
\providecommand{\url}[1]{\texttt{#1}}
\providecommand{\href}[2]{#2}
\providecommand{\path}[1]{#1}
\providecommand{\DOIprefix}{doi:}
\providecommand{\ArXivprefix}{arXiv:}
\providecommand{\URLprefix}{URL: }
\providecommand{\Pubmedprefix}{pmid:}
\providecommand{\doi}[1]{\href{http://dx.doi.org/#1}{\path{#1}}}
\providecommand{\Pubmed}[1]{\href{pmid:#1}{\path{#1}}}
\providecommand{\bibinfo}[2]{#2}
\ifx\xfnm\relax \def\xfnm[#1]{\unskip,\space#1}\fi
\bibitem[{Shtok et~al.(2012)Shtok, Kurland, Carmel, Raiber, and Markovits}]{NQC}
\bibinfo{author}{A.~Shtok}, \bibinfo{author}{O.~Kurland}, \bibinfo{author}{D.~Carmel}, \bibinfo{author}{F.~Raiber}, \bibinfo{author}{G.~Markovits},
\newblock \bibinfo{title}{Predicting query performance by query-drift estimation},
\newblock \bibinfo{journal}{ACM Trans. Inf. Syst.} \bibinfo{volume}{30} (\bibinfo{year}{2012}).
\bibitem[{Datta et~al.(2022)Datta, Ganguly, Mitra, and Greene}]{WRIG}
\bibinfo{author}{S.~Datta}, \bibinfo{author}{D.~Ganguly}, \bibinfo{author}{M.~Mitra}, \bibinfo{author}{D.~Greene},
\newblock \bibinfo{title}{A relative information gain-based query performance prediction framework with generated query variants},
\newblock \bibinfo{journal}{ACM Trans. Inf. Syst.} \bibinfo{volume}{41} (\bibinfo{year}{2022}).
\bibitem[{Shtok et~al.(2016)Shtok, Kurland, and Carmel}]{ReferenceBasedQPP}
\bibinfo{author}{A.~Shtok}, \bibinfo{author}{O.~Kurland}, \bibinfo{author}{D.~Carmel},
\newblock \bibinfo{title}{Query performance prediction using reference lists},
\newblock \bibinfo{journal}{ACM Trans. Inf. Syst.} \bibinfo{volume}{34} (\bibinfo{year}{2016}).
\bibitem[{Zendel et~al.(2019)Zendel, Shtok, Raiber, Kurland, and Culpepper}]{Oleg_2019}
\bibinfo{author}{O.~Zendel}, \bibinfo{author}{A.~Shtok}, \bibinfo{author}{F.~Raiber}, \bibinfo{author}{O.~Kurland}, \bibinfo{author}{J.~S. Culpepper},
\newblock \bibinfo{title}{Information needs, queries, and query performance prediction},
\newblock in: \bibinfo{booktitle}{proc. of {SIGIR'19}}, \bibinfo{year}{2019}, p. \bibinfo{pages}{395–404}.
\bibitem[{Gospodinov et~al.(2023)Gospodinov, MacAvaney, and Macdonald}]{Doc2Query--}
\bibinfo{author}{M.~Gospodinov}, \bibinfo{author}{S.~MacAvaney}, \bibinfo{author}{C.~Macdonald},
\newblock \bibinfo{title}{Doc2query--: When less is more},
\newblock in: \bibinfo{booktitle}{Advances in Information Retrieval}, \bibinfo{publisher}{Springer Nature Switzerland}, \bibinfo{address}{Cham}, \bibinfo{year}{2023}, pp. \bibinfo{pages}{414--422}.
\bibitem[{Lavrenko and Croft(2001)}]{Lavrenko_RLM2001:RBL:383952.383972}
\bibinfo{author}{V.~Lavrenko}, \bibinfo{author}{W.~B. Croft},
\newblock \bibinfo{title}{Relevance based language models},
\newblock in: \bibinfo{booktitle}{proc. of {SIGIR'01}}, \bibinfo{year}{2001}, p. \bibinfo{pages}{120–127}.
\bibitem[{Webber et~al.(2010)Webber, Moffat, and Zobel}]{rbo}
\bibinfo{author}{W.~Webber}, \bibinfo{author}{A.~Moffat}, \bibinfo{author}{J.~Zobel},
\newblock \bibinfo{title}{A similarity measure for indefinite rankings},
\newblock \bibinfo{journal}{ACM Trans. Inf. Syst.} \bibinfo{volume}{28} (\bibinfo{year}{2010}).
\bibitem[{Arabzadeh et~al.(2020)Arabzadeh, Zarrinkalam, Jovanovic, Al-Obeidat, and Bagheri}]{NeuralPreQPP}
\bibinfo{author}{N.~Arabzadeh}, \bibinfo{author}{F.~Zarrinkalam}, \bibinfo{author}{J.~Jovanovic}, \bibinfo{author}{F.~Al-Obeidat}, \bibinfo{author}{E.~Bagheri},
\newblock \bibinfo{title}{Neural embedding-based specificity metrics for pre-retrieval query performance prediction},
\newblock \bibinfo{journal}{{IPM}} \bibinfo{volume}{57} (\bibinfo{year}{2020}) \bibinfo{pages}{102248}.
\bibitem[{Roy et~al.(2019)Roy, Ganguly, Mitra, and Jones}]{embWVQPP}
\bibinfo{author}{D.~Roy}, \bibinfo{author}{D.~Ganguly}, \bibinfo{author}{M.~Mitra}, \bibinfo{author}{G.~J. Jones},
\newblock \bibinfo{title}{{Estimating Gaussian Mixture Models in the local neighbourhood of Embedded Word Vectors for Query Performance Prediction}},
\newblock \bibinfo{journal}{{IPM}} \bibinfo{volume}{56} (\bibinfo{year}{2019}) \bibinfo{pages}{1026--1045}.
\bibitem[{Cronen-Townsend et~al.(2002)Cronen-Townsend, Zhou, and Croft}]{Clarity}
\bibinfo{author}{S.~Cronen-Townsend}, \bibinfo{author}{Y.~Zhou}, \bibinfo{author}{W.~B. Croft},
\newblock \bibinfo{title}{Predicting query performance},
\newblock in: \bibinfo{booktitle}{proc. of {SIGIR'02}}, \bibinfo{year}{2002}, p. \bibinfo{pages}{299–306}.
\bibitem[{Cummins et~al.(2011)Cummins, Jose, and O'Riordan}]{ImproveQPPbyStandardDeviation}
\bibinfo{author}{R.~Cummins}, \bibinfo{author}{J.~Jose}, \bibinfo{author}{C.~O'Riordan},
\newblock \bibinfo{title}{Improved query performance prediction using standard deviation},
\newblock in: \bibinfo{booktitle}{proc. of {SIGIR'11}}, \bibinfo{publisher}{Association for Computing Machinery}, \bibinfo{year}{2011}, p. \bibinfo{pages}{1089–1090}.
\bibitem[{Shtok et~al.(2010)Shtok, Kurland, and Carmel}]{uef_kurland_sigir10}
\bibinfo{author}{A.~Shtok}, \bibinfo{author}{O.~Kurland}, \bibinfo{author}{D.~Carmel},
\newblock \bibinfo{title}{Using statistical decision theory and relevance models for query-performance prediction},
\newblock in: \bibinfo{booktitle}{Proceedings of the 33rd International ACM SIGIR Conference on Research and Development in Information Retrieval}, SIGIR '10, \bibinfo{publisher}{Association for Computing Machinery}, \bibinfo{year}{2010}, p. \bibinfo{pages}{259–266}.
\bibitem[{Jayasinghe et~al.(2015)Jayasinghe, Webber, Sanderson, Dharmasena, and Culpepper}]{statCompNonDetIRSysUTwoDimVar}
\bibinfo{author}{G.~K. Jayasinghe}, \bibinfo{author}{W.~Webber}, \bibinfo{author}{M.~Sanderson}, \bibinfo{author}{L.~S. Dharmasena}, \bibinfo{author}{J.~S. Culpepper},
\newblock \bibinfo{title}{Statistical comparisons of non-deterministic ir systems using two dimensional variance},
\newblock \bibinfo{journal}{{IPM}} \bibinfo{volume}{51} (\bibinfo{year}{2015}) \bibinfo{pages}{677--694}.
\bibitem[{Jaleel et~al.(2004)Jaleel, Allan, Croft, Diaz, Larkey, Li, Smucker, and Wade}]{NasreenJaleel_RM3}
\bibinfo{author}{N.~A. Jaleel}, \bibinfo{author}{J.~Allan}, \bibinfo{author}{W.~B. Croft}, \bibinfo{author}{F.~Diaz}, \bibinfo{author}{L.~S. Larkey}, \bibinfo{author}{X.~Li}, \bibinfo{author}{M.~D. Smucker}, \bibinfo{author}{C.~Wade},
\newblock \bibinfo{title}{Umass at {TREC} 2004: Novelty and {HARD}},
\newblock in: \bibinfo{booktitle}{TREC 2004}, \bibinfo{year}{2004}, pp. \bibinfo{pages}{1--13}.
\bibitem[{Mikolov et~al.(2013)Mikolov, Chen, Corrado, and Dean}]{w2v:mikolov}
\bibinfo{author}{T.~Mikolov}, \bibinfo{author}{K.~Chen}, \bibinfo{author}{G.~Corrado}, \bibinfo{author}{J.~Dean},
\newblock \bibinfo{title}{Efficient estimation of word representations in vector space},
\newblock in: \bibinfo{booktitle}{1st International Conference on Learning Representations, {ICLR} 2013, Scottsdale, Arizona, USA, May 2-4, 2013, Workshop Track Proceedings}, \bibinfo{year}{2013}, pp. \bibinfo{pages}{1--12}.
\bibitem[{Ebrahimi et~al.(2024)Ebrahimi, Khodabakhsh, Arabzadeh, and Bagheri}]{ContextRichQPP}
\bibinfo{author}{S.~Ebrahimi}, \bibinfo{author}{M.~Khodabakhsh}, \bibinfo{author}{N.~Arabzadeh}, \bibinfo{author}{E.~Bagheri},
\newblock \bibinfo{title}{Estimating query performance through rich contextualized query representations},
\newblock in: \bibinfo{booktitle}{Advances in Information Retrieval}, \bibinfo{publisher}{Springer Nature Switzerland}, \bibinfo{address}{Cham}, \bibinfo{year}{2024}, pp. \bibinfo{pages}{49--58}.
\bibitem[{Reimers and Gurevych(2019)}]{SBERT}
\bibinfo{author}{N.~Reimers}, \bibinfo{author}{I.~Gurevych},
\newblock \bibinfo{title}{Sentence-{BERT}: Sentence embeddings using {S}iamese {BERT}-networks},
\newblock in: \bibinfo{booktitle}{{proc. of EMNLP-IJCNLP}}, \bibinfo{year}{2019}, pp. \bibinfo{pages}{3982--3992}.
\bibitem[{Bajaj et~al.(2016)Bajaj, Campos, Craswell, Deng, Gao, Liu, Majumder, McNamara, Mitra, Nguyen et~al.}]{MSMARCO-DATASET}
\bibinfo{author}{P.~Bajaj}, \bibinfo{author}{D.~Campos}, \bibinfo{author}{N.~Craswell}, \bibinfo{author}{L.~Deng}, \bibinfo{author}{J.~Gao}, \bibinfo{author}{X.~Liu}, \bibinfo{author}{R.~Majumder}, \bibinfo{author}{A.~McNamara}, \bibinfo{author}{B.~Mitra}, \bibinfo{author}{T.~Nguyen}, et~al.,
\newblock \bibinfo{title}{{MS MARCO}: A human generated machine reading comprehension dataset},
\newblock \bibinfo{journal}{arXiv preprint arXiv:1611.09268}  (\bibinfo{year}{2016}).
\bibitem[{Craswell et~al.(2020)Craswell, Mitra, Yilmaz, Campos, and Voorhees}]{TREC-DL-2019-overview}
\bibinfo{author}{N.~Craswell}, \bibinfo{author}{B.~Mitra}, \bibinfo{author}{E.~Yilmaz}, \bibinfo{author}{D.~Campos}, \bibinfo{author}{E.~Voorhees},
\newblock \bibinfo{title}{{Overview of the TREC 2019 Deep Learning track}},
\newblock \bibinfo{journal}{CoRR} \bibinfo{volume}{abs/2003.07820} (\bibinfo{year}{2020}). \href{http://arxiv.org/abs/2003.07820}{{\tt arXiv:2003.07820}}.
\bibitem[{Craswell et~al.(2021)Craswell, Mitra, Yilmaz, and Campos}]{TREC-DL-2020-overview}
\bibinfo{author}{N.~Craswell}, \bibinfo{author}{B.~Mitra}, \bibinfo{author}{E.~Yilmaz}, \bibinfo{author}{D.~Campos},
\newblock \bibinfo{title}{{Overview of the TREC 2020 Deep Learning track}},
\newblock \bibinfo{journal}{CoRR} \bibinfo{volume}{abs/2003.07820} (\bibinfo{year}{2021}). \href{http://arxiv.org/abs/2102.07662}{{\tt arXiv:2102.07662}}.
\bibitem[{Nogueira et~al.(2020)Nogueira, Jiang, Pradeep, and Lin}]{monot5}
\bibinfo{author}{R.~F. Nogueira}, \bibinfo{author}{Z.~Jiang}, \bibinfo{author}{R.~Pradeep}, \bibinfo{author}{J.~Lin},
\newblock \bibinfo{title}{Document ranking with a pretrained sequence-to-sequence model},
\newblock in: \bibinfo{booktitle}{Findings of the Association for Computational Linguistics: {EMNLP} 2020, Online Event, 16-20 November 2020}, volume \bibinfo{volume}{{EMNLP} 2020} of \textit{\bibinfo{series}{Findings of {ACL}}}, \bibinfo{publisher}{Association for Computational Linguistics}, \bibinfo{year}{2020}, pp. \bibinfo{pages}{708--718}.
\bibitem[{Li et~al.(2023)Li, Yates, MacAvaney, He, and Sun}]{PARADE}
\bibinfo{author}{C.~Li}, \bibinfo{author}{A.~Yates}, \bibinfo{author}{S.~MacAvaney}, \bibinfo{author}{B.~He}, \bibinfo{author}{Y.~Sun},
\newblock \bibinfo{title}{{PARADE: Passage Representation Aggregation for Document Reranking}},
\newblock \bibinfo{journal}{ACM Trans. Inf. Syst.} \bibinfo{volume}{42} (\bibinfo{year}{2023}).
\bibitem[{Devlin et~al.(2019)Devlin, Chang, Lee, and Toutanova}]{BERT}
\bibinfo{author}{J.~Devlin}, \bibinfo{author}{M.~Chang}, \bibinfo{author}{K.~Lee}, \bibinfo{author}{K.~Toutanova},
\newblock \bibinfo{title}{{BERT:} pre-training of deep bidirectional transformers for language understanding},
\newblock in: \bibinfo{booktitle}{proc. of {NAACL-HLT'19}}, \bibinfo{year}{2019}, pp. \bibinfo{pages}{4171--4186}.
\bibitem[{Lin et~al.(2021)Lin, Yang, and Lin}]{TCT-ColBERT}
\bibinfo{author}{S.-C. Lin}, \bibinfo{author}{J.-H. Yang}, \bibinfo{author}{J.~Lin},
\newblock \bibinfo{title}{In-batch negatives for knowledge distillation with tightly-coupled teachers for dense retrieval},
\newblock in: \bibinfo{booktitle}{Proceedings of the 6th Workshop on Representation Learning for NLP (RepL4NLP-2021)}, \bibinfo{year}{2021}, pp. \bibinfo{pages}{163--173}.
\bibitem[{Khattab and Zaharia(2020)}]{ColBERT}
\bibinfo{author}{O.~Khattab}, \bibinfo{author}{M.~Zaharia},
\newblock \bibinfo{title}{{ColBERT: Efficient and Effective Passage Search via Contextualized Late Interaction over BERT}},
\newblock in: \bibinfo{booktitle}{Proceedings of the 43rd International ACM SIGIR Conference on Research and Development in Information Retrieval}, SIGIR '20, \bibinfo{year}{2020}, p. \bibinfo{pages}{39–48}.
\bibitem[{Ganguly et~al.(2022)Ganguly, Datta, Mitra, and Greene}]{AnalyseVariantInQPP}
\bibinfo{author}{D.~Ganguly}, \bibinfo{author}{S.~Datta}, \bibinfo{author}{M.~Mitra}, \bibinfo{author}{D.~Greene},
\newblock \bibinfo{title}{An analysis of variations in the effectiveness of query performance prediction},
\newblock \bibinfo{journal}{CoRR} \bibinfo{volume}{abs/2202.06306} (\bibinfo{year}{2022}).
\bibitem[{Arabzadeh et~al.(2021)Arabzadeh, Khodabakhsh, and Bagheri}]{qppbert}
\bibinfo{author}{N.~Arabzadeh}, \bibinfo{author}{M.~Khodabakhsh}, \bibinfo{author}{E.~Bagheri},
\newblock \bibinfo{title}{{BERT-QPP:} contextualized pre-trained transformers for query performance prediction},
\newblock in: \bibinfo{booktitle}{{CIKM}}, \bibinfo{publisher}{{ACM}}, \bibinfo{year}{2021}, pp. \bibinfo{pages}{2857--2861}.
\bibitem[{Macdonald et~al.(2021)Macdonald, Tonellotto, MacAvaney, and Ounis}]{pyterrier}
\bibinfo{author}{C.~Macdonald}, \bibinfo{author}{N.~Tonellotto}, \bibinfo{author}{S.~MacAvaney}, \bibinfo{author}{I.~Ounis},
\newblock \bibinfo{title}{{PyTerrier}: Declarative experimentation in python from bm25 to dense retrieval},
\newblock in: \bibinfo{booktitle}{proc. of {CIKM '21}}, \bibinfo{year}{2021}, p. \bibinfo{pages}{4526–4533}.

\end{thebibliography}
